\documentclass[useAMS]{mn2e}

\usepackage{epsfig,amsmath,amssymb,amsfonts,mathrsfs,latexsym,graphicx,lscape}

\begin{document}

\title[Line and disk emission in Swift J1753.5$-$0127]{Discovery of a broad iron line in the black-hole candidate Swift J1753.5$-$0127, and the disk emission in the low/hard state revisited.}

\author[B. Hiemstra et al.]
       {Beike Hiemstra$^{1}$\thanks{E-mail: hiemstra@astro.rug.nl}, Paolo Soleri$^{2}$, Mariano M\'{e}ndez$^{1}$, Tomaso Belloni$^{3}$,
\newauthor Reham Mostafa$^{4}$ and Rudy Wijnands$^{2}$\\
        $^{1}$Kapteyn Astronomical Institute, University of Groningen, P.O. Box 800, 9700 AV Groningen, The Netherlands\\
        $^{2}$Astronomical Institute Anton Pannekoek, University of Amsterdam, Kruislaan 403, 1098 SJ Amsterdam, The Netherlands\\
	$^{3}$INAF -- Osservatorio Astronomico di Brera, Via E. Bianchi 46, I-23807, Merate (LC), Italy\\
        $^{4}$Department of Physics, Fayoum University, P.O. Box 63514, Fayoum, Egypt}
\date{      }

\pagerange{\pageref{firstpage}--\pageref{lastpage}}
\pubyear{2008}

\maketitle

\label{firstpage}

\begin{abstract}
We analyzed simultaneous archival \textit{XMM-Newton} and \textit{RXTE} observations of the X-ray binary and black hole candidate Swift J1753.5$-$0127. In a previous analysis of the same data a soft thermal component was found in the X-ray spectrum, and the presence of an accretion disk extending close to the innermost stable circular orbit was proposed. This is in contrast with the standard picture in which the accretion disk is truncated at large radii in the low/hard state. We tested a number of spectral models and we found that several of them fit the observed spectra without the need of a soft disk-like component. This result implies that the classical paradigm of a truncated accretion disk in the low/hard state can not be ruled out by these data. We further discovered a broad iron emission line between 6 and 7 keV in these data. From fits to the line profile we found an inner disk radius that ranges between $\sim$6--16 gravitational radii, which can be in fact much larger, up to $\sim$250 gravitational radii, depending on the model used to fit the continuum and the line. We discuss the implications of these results in the context of a fully or partially truncated accretion disk.
\end{abstract}


\section{Introduction}
Swift J1753.5$-$0127 (hereafter Swift J1753) was discovered with the \textit{Swift} Burst Alert Telescope on June 30 2005 as a hard X-ray source ($E > 15$ keV; Palmer et al. 2005). Swift J1753 was also detected at energies below 10 keV with the \textit{Swift} X-Ray Telescope (XRT), and in the UV with the \textit{Swift} UV/Optical Telescope (Morris et al. 2005; Still et al. 2005). Using optical spectroscopy, Torres et al. (2005) found double peaked H$\alpha$ emission lines, which indicate that the system is a low-mass X-ray binary. The source was also detected in radio, which might evidence jet activity (Fender et al. 2005; Cadolle Bel et al. 2006). Based on the hard spectral behavior, Cadolle Bel et al. (2005) proposed that Swift J1753 is a black hole candidate. Although, it is generally assumed that Swift J1753 contains a black hole, there is yet no dynamical confirmation of its nature. Bearing this in mind, in the rest of this paper we treat Swift J1753 as a black hole candidate (BHC).

\textit{Rossi X-ray Timing Explorer} (\textit{RXTE}) and \textit{INTEGRAL} observations showed that Swift J1753 never left the low/hard state (LHS) during the whole outburst (Cadolle Bel et al. 2007; Zhang et al. 2007). The LHS of X-ray binaries is characterized by a hard power-law X-ray spectrum (photon index, $\Gamma$$\sim$1.7; e.g., M\'{e}ndez \& van der Klis 1997; Remillard \& McClintock 2006), which is usually interpreted as the result of Comptonization of thermal photons from an optically thick accretion disk by hot electrons (e.g., Esin et al. 1997). Observations of sources with low absorption provide evidence that in the LHS the accretion disk is cool and is truncated at a large radius (e.g., McClintock et al. 2001). In the last years, a contribution of a jet to the X-ray emission in the LHS has been considered (Markoff et al. 2001; Fender, Belloni \& Gallo 2004). 

During an outburst, the mass accretion rate onto the black hole increases and a transition to the high/soft state (HSS) can occur (for reviews, see Homan \& Belloni 2005; McClintock \& Remillard 2006). The high/soft state is dominated by thermal emission below $\sim$5 keV, likely produced by an optically thick and geometrically thin accretion disk with an inner radius at, or close to, the innermost stable circular orbit (ISCO; Shakura \& Sunyaev 1973). Due to the increase of thermal photons, the Comptonizing plasma in the system cools down and the spectrum of the source becomes soft. A part of the Comptonized photons is reflected off the accretion disk, and produces an iron K$\alpha$ line between 6 and 7 keV (e.g., Fabian et al. 1989; Laor 1991), where the width of the emission line is related to the size of the inner disk radius. In this interpretation, if the disk is truncated at large radii in the LHS, no or narrow iron-emission lines are expected, whereas the iron line would be broad in the HSS where the inner disk radius is close to the ISCO and the relativistic effects are strong.

Recent high-resolution spectral data of the BHC X-ray binaries Swift J1753, GX 339$-$4 and XTE J1817$-$330, showed a soft thermal component in the spectra while the sources were in the LHS (Miller et al. 2006a, hereafter M06; Miller et al. 2006b; Rykoff et al. 2007). It was suggested that the soft component is caused by thermal emission from an accretion disk which, even in the LHS, extends down to the ISCO. The same data set of XTE J1817$-$330 was re-analyzed by Gierli\'{n}ski et al. (2008), who showed that the spectrum could also be fit without the need of a thermal component.

Here we analyzed a simultaneous archival \textit{XMM-Newton}/\textit{RXTE} (the same data as reported in M06) observation of Swift J1753, and we tested whether a thermal disk-like component is required to fit the X-ray spectra. We explored a range of continuum models, beyond the models described in M06, with and without a disk component. We also report on the presence of a broad Fe line in the spectra. In \S\ref{sec:reduction} we describe the extraction and reduction of the data, in \S\ref{sec:data analysis} we present the results of fitting the data with different continuum models, and in \S\ref{sec:discussion} we discuss the results of the continuum models and the line components with respect to the issue of a radially truncated accretion disk.
\section{Data reduction}\label{sec:reduction}
\setcounter{figure}{0}
\begin{figure*}
\begin{center}
\resizebox{2\columnwidth}{!}{\rotatebox{-90}{\includegraphics[clip]{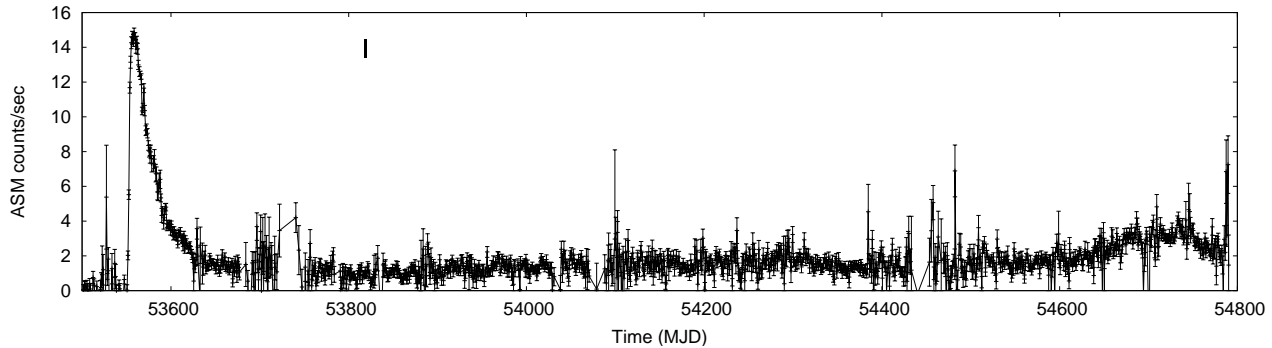}}}
\caption{\textit{RXTE}/ASM light curve of Swift J1753 with daily points, starting at the outburst in 2005 (MJD 53551). The \textit{XMM-Newton/RXTE} observation, $\sim$270 days after the start of the outburst, is represented by the vertical bar. Note that 3 years after the start of the outburst, the source is still active.}
\label{fig:lightcurve}
\end{center}
\end{figure*}
We analyzed archival observations of Swift J1753 simultaneously taken with \textit{XMM-Newton} and \textit{RXTE} on March 24, 2006, i.e. $\sim$270 days after the start of the outburst. \textit{XMM-Newton} started its $\sim$42 ks observation at 16:00:31 (UT), and the \textit{RXTE} observation started at 17:32:16 (UT). In Fig.~\ref{fig:lightcurve} we show the date of this observation on top of the All Sky Monitor (ASM) \textit{RXTE} light curve of Swift J1753, starting from its outburst in 2005. In the next subsections we describe the procedure that we followed to extract the spectra for each of the different telescopes.
\subsection{\textit{XMM-Newton}}
We reduced the \textit{XMM-Newton} Observation Data Files (ODF) using SAS version 7.1.2 and we applied the latest calibration files. We extracted the event files for the EPIC-pn (hereafter PN) and both EPIC-MOS cameras (MOS1 and MOS2) using the tasks \texttt{epproc} and \texttt{emproc}, respectively. We applied standard filtering and examined the light curves for background flares. No flares are present and we used the whole exposure for our analysis. We checked the filtered event files for photon pile-up by running the task \texttt{epatplot}. No pile-up is apparent in the PN and MOS1 data since both cameras were operated in `timing' mode. The MOS1 data, however, show a sharp drop in the histogram in which counts versus RAWX\footnote{In timing mode, one of the spatial coordinates of the CCD is compressed in order to increase the time resolution. For \textit{XMM-Newton} the spatial coordinate that is not compressed in this mode is called RAWX.} are plotted. This feature is due to a physical damage of some pixels in the MOS1 camera as a consequence of a micro-meteorite impact. For this reason, we excluded the MOS1 data from our analysis. The MOS2 camera was operated in `full-frame' mode and suffered from moderate photon pile-up. We therefore excluded the central 22$''$ region of the source point spread function (PSF) and confirmed, using the task \texttt{epatplot}, that the remaining MOS2 data are not affected by pile-up. The ODF files also include high resolution spectra at energies below 2 keV, taken with the Reflection Grating Spectrometer (RGS). Since the PN and MOS cameras cover most of the range of the RGS spectra, we do not use the RGS data in this paper.

We extracted source and background spectra applying the standard procedures. The source spectrum for the PN data is extracted using RAWX in [28:48], and the background in [5:25]. For MOS2, we generated the source spectrum using an annulus with inner radius of 22$''$ and outer radius of 68$''$ centered on the source location. We extracted the MOS2 background spectrum from a circular region with a radius of 160$''$ located 10.5$'$ away from the position of the source. We created photon redistribution matrices and ancillary files for the source spectra using the SAS tools \texttt{rmfgen} and \texttt{arfgen}, respectively.

We rebinned the source spectra using the tool \texttt{pharbn}\footnote{The \texttt{pharbn} tool allows to rebin the energy channels of the spectra for different instruments taking into account the instrument's spectral resolution. It allows to rebin both on minimum number of bins per resolution element and on minimum number of counts per bin} (M. Guainazzi, private communication), such that the number of bins per resolution element of the PN and MOS2 spectra is 3. We checked that after rebinning, all channels have at least 15 counts in the energy range used in our fits. We note that differently from us, M06 grouped the PN data to have at least 10 counts per bin, and they therefore oversampled the intrinsic resolution of the PN data by a factor of $\sim$8. The cross calibration between PN and MOS agrees to 7\% from 0.4 keV to 12 keV\footnote{http://xmm2.esac.esa.int/docs/documents/CAL-TN-0018.pdf}, and we found that only below 2 keV the systematic difference in the relative effective areas of these cameras went up to 7\%. We therefore added a 7\% systematical error to the PN and MOS2 spectra below 2 keV.
\subsection{\textit{RXTE}}
We reduced the \textit{RXTE} Proportional Counter Array (PCA) and High Energy X-ray Timing Experiment (HEXTE) data using the HEASOFT tools version 6.4. We applied standard filtering and obtained PCA and HEXTE exposures which are $\sim$2.3 ks and $\sim$0.8 ks, respectively. Using the tool \texttt{saextrct}, we extracted PCA spectra from PCU2 only, being this the best calibrated detector and the only one which is always on. For the HEXTE data we generated the spectra using cluster-B events only, since after January 2006 cluster A stopped rocking and could no longer measure the background. We estimated the PCA background and measured the HEXTE background using the standard \textit{RXTE} tools \texttt{pcabackest} and \texttt{hxtback}, respectively. We further built instrument response files for the PCA and HEXTE data using \texttt{pcarsp} and \texttt{hxtrsp}, respectively. The energy channels of the \textit{RXTE} spectra have more than 20 counts per energy bin so that $\chi^{2}$ statistics can be applied, therefore no channel rebinning is required. Following M06, we added a 0.6\% systematic error to the PCA spectra using the FTOOL \texttt{grppha}, and we did not add any systematic error to the HEXTE data.
\section{Data analysis and results}\label{sec:data analysis}
For our spectral analysis of Swift J1753 we used XSPEC v11 (Arnaud 1996). We simultaneously fit the spectra of the different instruments, where we restricted the PN and the MOS2 spectra to an energy range of 0.6--10 keV and 0.6--9.4 keV, respectively. For the PCA and HEXTE spectra we used the energy ranges 3--20 keV and 20--100 keV, respectively. To investigate the nature of the putative disk emission in the LHS (as reported in M06), we tested several models to describe the continuum spectrum of Swift J1753.
\subsection{General fit procedure}\label{sec:general fit}
Each of the spectral continuum models we present below includes the effect of interstellar absorption in the direction of Swift J1753, using the \textsc{phabs} component with cross-sections from Ba{\l}uci\'{n}ska-Church \& McCammon (1992) and abundances from Anders \& Grevesse (1989). The total column density in the direction of the source, as derived from HI data, is $N_{\rm H} = 1.7 \times 10^{21}$ cm$^{-2}$ (Dickey \& Lockman, 1990).  Since this column density provides an upper limit to the interstellar absorption to Swift J1753, we left $N_{\rm H}$ free to vary during our fits. For an overall normalization between the different instruments, we multiplied the continuum model by a \textsc{constant} component. All model parameters, except these multiplicative constants, were linked between the different instruments.

In the fit residuals of the PN and MOS2 data, two narrow absorption lines near 1.8 keV and 2.2 keV were present. This is most likely due to a mismatch in the calibration of the edges in the EPIC instruments. To reduce the impact of these features in our fit results, we fit Gaussian absorption lines, which we found to be at an energy of 1.82$\pm$0.01 keV and 2.25$\pm$0.03 keV, with widths of 0.23$\pm$0.15 eV and 2.0$\pm$1.9 eV, respectively.

There appear to be a broad emission line between 6 and 7 keV in the \textit{XMM-Newton/RXTE} spectra. (This emission line is best seen in the PN spectra; MOS2 effective area is a factor of $\sim$2 smaller than that of PN, and in this case the effective area has been reduced due to the excision of the central part of the source PSF to avoid pile-up. On the other hand, the spectral resolution of \textit{RXTE} is $\sim$1 keV at 6 keV.) Similar emission lines, probably due to iron, have been reported in other BHC (e.g., Miller et al. 2002b) and active galactic nuclei (e.g., Tanaka et al. 1995). We initially added a Gaussian component with a variable line width to the spectrum to fit this line, but there always remained residuals around 7 keV which we could only eliminate by adding at least another Gaussian to our model. On the contrary, a \textsc{Laor} profile (Laor 1991) yields a good fit of the line and therefore, in the rest of the paper we use this relativistic line profile whenever we fit a line to the spectra, although we also tried different relativistic line profiles as described in \S\ref{sec:line}. For the \textsc{Laor} component we allowed all the parameters to be free to vary, except the outer disk radius, which we fixed to the default value of 400 R$_{\rm g}$ (R$_{\rm g}$ is the gravitational radius of the black hole, defined as $R_{g} = G M/c^{2}$, with $G$ and $c$ the common physical constants, and $M$ the mass of the black hole). Further, we constrained the line energy to the range between 6.4 keV and 7 keV, which is the energy range spanned by the K$\alpha$ lines from neutral to hydrogen-like iron.

For all fits of the different continuum models, the values of the parameters and the 90\% confidence errors are given in Tab.~\ref{tab:more models}. In the text we only give approximate numbers of the parameters.
\subsection{A power-law model}\label{sec:pl}
The first continuum model we fit was a power-law model (\textsc{pl}). This model yields a poor fit ($\chi^{2}/\nu = 845/502$). Following M06, we added a disk blackbody component \textsc{diskbb} (Mitsuda et al. 1984) with a best-fit inner disk temperature of $\sim$0.4 keV; the reduced $\chi^{2}$ in this case is 1.21 ($\chi^{2}/\nu$ are given in Tab.~\ref{tab:more models}). The spectra and fit residuals of the \textsc{pl+diskbb} model are shown in the left panel of Fig.~\ref{fig:diskbb+pl}. A broad emission line between 6--7 keV is apparent in this figure. We added a \textsc{Laor} line profile to the \textsc{pl+diskbb} model and the fit improved significantly, with a reduced $\chi^{2}$ of 1.14. To check whether a fit to the spectrum with only a line and a power-law component provides a good fit, we fit a \textsc{pl+Laor} model, without the \textsc{diskbb} component. The \textsc{pl+Laor} model gives a reduced $\chi^{2}$ of 1.30. In the best-fit model (i.e. the \textsc{pl+Laor+diskbb} model), the line is required at high confidence as indicated by the fact that the normalization of the line is 10$\sigma$ different from zero. The line in the best-fit model has an equivalent width of 90.3 eV, and the \textsc{diskbb} component is required at a confidence level of more than 8$\sigma$. The flux of the \textsc{diskbb} component is 7.2$\times$10$^{-12}$ erg cm$^{-2}$ s$^{-1}$ (0.6--10 keV), where the unabsorbed flux of the best-fit model is 3.9$\times 10^{-10}$ erg cm$^{-2}$ s$^{-1}$ (0.6--10 keV). We note that the \textsc{diskbb} normalization that we obtained is much smaller than the value that M06 found, and the contribution from the disk to the total flux that we found is a factor of $\sim$3 less than reported in M06.

In the case when the distance, inclination angle of the disk, and mass of the black hole are known, the inner disk radius, $R_{\rm in}$ can be obtained from the \textsc{diskbb} normalization ($N_{\rm D}$). However, for the case of Swift J1753 these parameters are currently not known. Assuming that the source is closer than 10 kpc and has an inclination lower than 63$^{\circ}$, since eclipses are not yet reported, $R_{\rm in} \lesssim$ 1.48 $\sqrt{N_{\rm D}}$ km. Assuming a black hole mass $>$3 M$_{\odot}$, $R_{\rm in} <$0.33 $\sqrt{N_{\rm D}}$ R$_{\rm g}$. From the \textsc{diskbb} normalization we found a non-truncated disk, while inferred from the line profile we found $R_{\rm in}$ $\sim$15 R$_{\rm g}$. For an inclination angle $>$ 85$^{\circ}$, as determined from the line profile, $R_{\rm in}$ can become much larger.
\setcounter{figure}{1}
\begin{figure*}
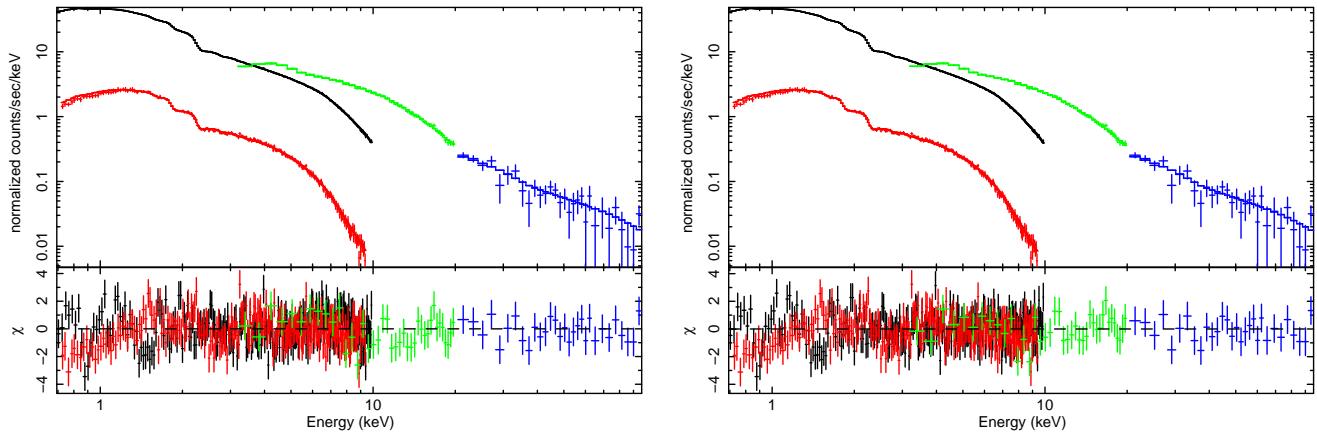

\resizebox{1\columnwidth}{!}{\rotatebox{-90}{\includegraphics[clip]{./Pl-Diskbb.ps}}}
\hspace{0.3cm}
\resizebox{1\columnwidth}{!}{\rotatebox{-90}{\includegraphics[clip]{./Diskbb-Pl-Laor.ps}}}
\caption{\textit{XMM-Newton} PN (0.3--10 keV), MOS2 (0.3--9.4 keV), \textit{RXTE} PCA (3--20 keV), and HEXTE (20--100 keV) spectra of Swift J1753. In the lower panels the best fit residuals of a \textsc{pl+diskbb} model (left) and a \textsc{pl+Laor+diskbb} model (right) are shown. The Laor line profile is fitted to the broad emission line that is apparent in the spectra at around 7 keV.}
\label{fig:diskbb+pl}
\end{figure*}
\subsection{Cut-off power-law model}\label{sec:cutoffpl}
Next, we added a high-energy cut off to the power-law component. In the case when we let all the parameters free, the high-energy cut off became larger than 200 keV. Since this value is outside the energy range of the spectra, this component is effectively the same as the simple power law. We therefore did not explore this model further.
\subsection{Broken power-law model}\label{sec:bknpl}
The next continuum model we tried is based on a broken power law (\textsc{bknpl}) instead of a power law. The fit of the \textsc{bknpl} model gives already a reasonable fit with a reduced $\chi^{2}$ of 1.20, which is similar to the value for the fit with a \textsc{pl+diskbb} model. The break energy is at 2.9 keV, with a photon index of 1.7 before the break, and 1.6 after the break. The latter photon index is similar to the photon index of the power-law component in the \textsc{pl+diskbb} model. 

A \textsc{Laor} component added to this continuum model gives a reduced $\chi^{2}$ of 1.14, where the break energy and photon indices are consistent with those in the fit without the line. We note that the \textsc{bknpl+Laor} model fits the spectra equally well as the \textsc{pl+Laor+diskbb} model does, with the parameters of the \textsc{Laor} component being consistent between the two models. The normalization of the line is more than 3$\sigma$ different from zero.

To see whether a soft thermal component is required, we included a \textsc{diskbb} component to the model. The fit with a \textsc{bknpl+Laor+diskbb} model gives a reduced $\chi^{2}$ of 1.12, which is a marginal improvement of the fit without the \textsc{diskbb} component. In contrast with the \textsc{Laor} component, the \textsc{diskbb} component is not significant, with a normalization that is $<$ 3$\sigma$ different from zero. The normalization of the line is 6$\sigma$ different from zero for the \textsc{bknpl+Laor+diskbb} model. The line in the best-fit model (i.e. \textsc{bknpl+Laor}) has an equivalent width of 67.5 eV.

We note that the part of the power law before the break at $\sim$3 keV possibly affects the contribution of the disk blackbody, and that therefore the \textsc{diskbb} component in the broken power-law model is not as significant compared with the continuum model based on a power law without a break.
\subsection{Comptonization model}\label{sec:comptt}
We then replaced the \textsc{bknpl} component with a model that describes the hard emission as Comptonization of soft photons in a hot plasma (\textsc{comptt}; Titarchuk 1994). Using a disk geometry for the Comptonizing material, the \textsc{comptt} model gives a poor fit ($\chi^{2}/\nu =$ 753/500). A spherical geometry does not provide a good fit either. Adding a \textsc{Laor} component gives a reduced $\chi^{2}$ of 1.34, with the normalization of the line being more than 7$\sigma$ different from zero.

The \textsc{comptt+Laor+diskbb} model gives a reduced $\chi^{2}$ of 1.15, with an inner disk temperature of $\sim$0.4 keV and a \textsc{diskbb} normalization that is $>$ 3$\sigma$ different from zero. The line in the best-fit model has an equivalent width of 59.9 eV, with a line normalization that is $>$ 4$\sigma$ different from zero. We note that the seed photon temperature of the \textsc{comptt} component becomes more than 10 times smaller than in the case when no \textsc{diskbb} is added to the continuum. The 0.6--10 keV flux of this three-component model is 3.9$\times$10$^{-10}$ erg cm$^{-2}$ s$^{-1}$, whereas the flux of the \textsc{diskbb} component is 8.9$\times$10$^{-12}$ erg cm$^{-2}$ s$^{-1}$.
\subsection{Reflection models: PEXRAV and PEXRIV}\label{sec:pexrav}
The presence of a broadened Fe-emission line between 6 and 7 keV suggests that the X-ray emission may partly be due to reflection of hard X-rays by a relatively cool disk. We therefore explored fits to the data that include this reflection. We first used the \textsc{pexrav} and \textsc{pexriv} models, which describe the Compton reflection by neutral and ionized material, respectively (Magdziarz \& Zdziarski 1995).

The incident spectrum is assumed to be a power law with a high-energy cut off. Initially, we allowed this cut-off energy to be free to vary, but it always converged to values much higher than 200 keV, outside the energy range covered by our instruments, and we therefore used no cut off. Further, we fixed all abundances to solar and the reflection factor to~1. Compared to \textsc{pexrav}, the \textsc{pexriv} model has two extra parameters: The ionization parameter $\xi$, which we left free to vary, and the disk temperature, $kT_{\rm disk}$, which we fixed to the default value of 3$\times 10^{4}$ K. 

The fit with the \textsc{pexrav} model gives a reduced $\chi^{2}$ of 1.38, and with the \textsc{pexriv} model the reduced $\chi^{2}$ = 1.30. The \textsc{pexrav/pexriv} models include only the effect of bound-free transitions in the reflected spectrum. To account for the iron emission line at around 7 keV (caused by bound-bound transitions), we added a separate \textsc{Laor} component to the reflection models. The spectral fits improved significantly, with a reduced $\chi^{2}$ of 1.19 and 1.18 for the \textsc{pexriv+Laor} and \textsc{pexrav+Laor} models, respectively. The normalization of the line is 9$\sigma$ different from zero in the fit with \textsc{pexrav}, and 8$\sigma$ different from zero in the fit with \textsc{pexriv}. The equivalent width of the line is 181 eV and 187 eV in the \textsc{pexrav+Laor} and \textsc{pexriv+Laor} models, respectively.

Adding a \textsc{diskbb} component to the reflection plus line model gives a reduced $\chi^{2}$ of 1.14 for both models, with a best-fit disk temperature $kT_{\rm in}$ = 0.3 keV and 0.4 keV for the neutral and ionized models, respectively. The \textsc{diskbb} component is statistically not required in the \textsc{pexrav+Laor+diskbb} model, with a \textsc{diskbb} normalization that is $<$ 3$\sigma$ different from zero. In contrast, for the \textsc{pexriv+Laor+diskbb} model the \textsc{diskbb} normalization is 7$\sigma$ different from zero. The flux of the \textsc{diskbb} component is 5.1$\times$ 10$^{-12}$ erg cm$^{-2}$ s$^{-1}$ and 5.5$\times$ 10$^{-12}$ erg cm$^{-2}$ s$^{-1}$ (0.6--10 keV) for the neutral and ionized reflection models, respectively.
\subsection{Ionized reflection model: REFLION}\label{sec:reflion}
Since \textsc{pexrav} and \textsc{pexriv} do not include the contribution of bound-bound transitions to the reflected spectrum, we also fit the data with a model that describes the reflection by an ionized, optically thick, illuminated atmosphere of constant density (\textsc{reflion}; Ross \& Fabian 2005). The atmosphere is illuminated by a power-law spectrum with an exponential cut off at high energy (the cut-off energy is fixed at 300 keV, and hence outside our energy range). This model includes ionization states and transitions for the most important ions at energies above 1~eV, like O III--VIII, Fe VI--XXVI, and several others. We fixed the iron abundance to solar. \textsc{Reflion} only provides the reflected emission, not the direct component, and we therefore used the \textsc{reflion+pl} model to fit the spectra, where we coupled the photon index of the illuminating power-law component in \textsc{reflion} to the photon index of the \textsc{pl}. 

The \textsc{reflion+pl} model gives a reduced $\chi^{2}$ of 1.14. In this case we do not need to include a component to fit the line, since the line is part of the reflected emission. The unabsorbed flux of this two-component model is 3.9$\times$10$^{-10}$~erg~cm$^{-2}$~s$^{-1}$ (0.6--10 keV), where the flux of the power-law component is 3.0$\times$10$^{-10}$~erg~cm$^{-2}$~s$^{-1}$ (0.6--10 keV). The reflection ratio in the 0.6--10 keV energy range is 0.3. Statistically, a \textsc{diskbb} component is not required to fit the data; adding this component gives a reduced $\chi^{2}$ of 1.13 with a best-fit temperature $kT_{\rm in}$ = 0.4 keV and a \textsc{diskbb} normalization which is less than 3$\sigma$ different from zero. The 95\% confidence limits for the \textsc{diskbb} component implies a disk flux $<$ 6.1$\times 10^{-12}$~erg~cm$^{-2}$~s$^{-1}$ (0.6--10 keV).

To gain information about the disk radius, we convolved \textsc{reflion} with the \textsc{Kerrconv} component (Brenneman \& Reynolds 2005). The \textsc{Kerrconv} kernel is applied to the reflected component to account for relativistic effects in the disk. We fixed the two emissivity indices to 3 (i.e. similar to the index that we found in the line profile) and the break radius to a large value (this is equivalent to a single emissivity law for the whole disk). We further fixed the outer radius of the disk to its maximum value. We found that the spin of the black hole is not constrained by these data, and since we are testing for the possibility of a truncated disk at a distance that is much larger than the radius of the ISCO, we fixed the spin to its maximum value of 0.998 and left the inner radius of the disk free to vary. This allows the disk radius to float from $\sim$1.23 R$_{\rm g}$ to its maximum value. We checked that fixing the spin to zero did not change the results significantly. In summary, we allowed only two parameters of \textsc{Kerrconv} to be free: The inner radius of the disk, and the inclination of the disk with respect to the line of sight. The \textsc{(reflion+pl)$\ast$Kerrconv} gives a reduced $\chi^{2}$ of 1.13, with an inner disk radius of 256 R$_{\rm g}$, an inclination angle consistent with 0$^{\circ}$, and the rest of the parameters consistent with the model parameters without the \textsc{Kerrconv} kernel. 
\subsection{The broad iron line at around 7 keV}\label{sec:line}
\setcounter{figure}{2}
\begin{figure*}
\resizebox{1\columnwidth}{!}{\rotatebox{-90}{\includegraphics[clip]{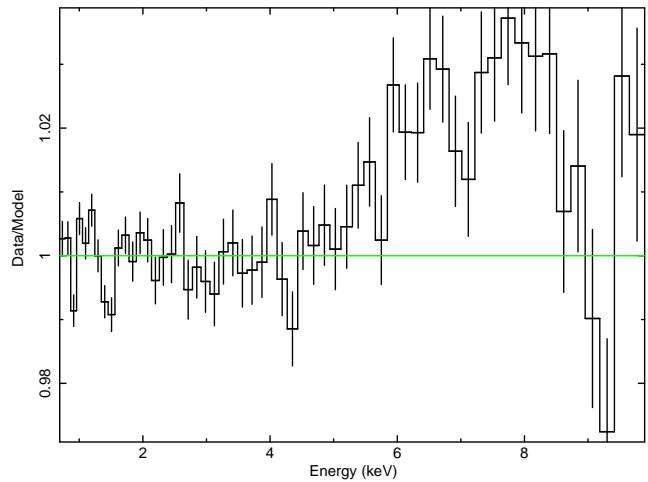}}}
\caption{In the various continuum models fit to the \textit{XMM-Newton/RXTE} spectra of Swift J1753, a broad emission line is apparent between 6 and 7 keV. Here, we show a zoom in of the PN residuals in the fit with the \textsc{pl+Laor+diskbb} model. To show the broad emission line, we have set the normalization of the line to zero, and applied an additional rebinning to the plot.}
\label{fig:residuals}
\end{figure*}
In previous subsections we included a \textsc{Laor} component to the continuum model to fit a broad line at around $\sim$7 keV. The line is significantly required by the fit, as indicated by the fact that the normalization of the line is different from zero at a confidence level of 3--14$\sigma$, depending on the model we used to fit the continuum. As shown in the right panel of Fig.~\ref{fig:residuals}, the line is extending from $\sim$5 keV up to $\sim$9 keV, with an equivalent width of $\sim$60--187 eV, depending on the continuum model used to fit the spectrum.

The best-fit parameters of the line depend on the model that we used to represent the underlying continuum emission. For instance, for the different best-fit models in Tab.~\ref{tab:more models}, the inner disk radius varies from 5.5 R$_{\rm g}$ to 14.6 R$_{\rm g}$ and the inclination appears to be constrained to high values, $i > 85^{\circ}$. Such high inclination seems to be incompatible with the fact that no eclipses have been observed in the X-ray light curve of Swift J1753. A possible explanation for this high inclination is that the broad line that we observe is a combination of lines from more than one ionization stage of iron. We tested this by adding three \textsc{Laor} profiles to the \textsc{pl+diskbb} model as described in \S\ref{sec:pl}, with the energy of each line fixed at 6.4 keV, 6.7 keV and 6.97 keV, respectively, to account for possible emission from neutral and lowly ionized, He-like, and H-like iron, respectively. We further coupled all the remaining parameters of these three lines, with only the line normalizations left free to vary. This multiple line model gives a reduced $\chi^{2}$ of 1.14, with the 95\% confidence limits of the inclination angle $83^{\circ}$ and $86^{\circ}$, and the line normalizations of the 6.4 keV and 6.7 keV lines being consistent with zero. A more realistic approach would be to link the relative normalizations of the iron lines such that their ratios would be determined by the ionization balance in the disk. We note that the \textsc{reflion} component, as described in \S\ref{sec:reflion}, calculates self-consistently the relative intensities of lines from different ionization stages of all elements in the reflected spectrum by fitting the ionization parameter of the disk. Recall that the fits of the \textsc{reflion} component convolved with \textsc{Kerrconv} in \S\ref{sec:reflion} yield a large inner disk radius ($\sim$250 R$_{\rm g}$) and a low inclination angle ($i < 8^{\circ}$ at 95\% confidence level).

The exact values of the line parameters may also depend on the approximations done to calculate the relativistic profile of the line. To test this, we tried three other components, besides the \textsc{Laor} profile, to represent the line profile: \textsc{diskline} (Fabian et al. 1989), \textsc{Kerrdisk} (Brenneman \& Reynolds 2005), and \textsc{Kyrline} (Dov{\v c}iak, Karas \& Yaqoob 2004). We only applied these components to the \textsc{pl+diskbb} model, such that we either fit \textsc{pl+diskbb+diskline}, \textsc{pl+diskbb+Kerrdisk} or \textsc{pl+diskbb+Kyrline}.  For all 3 models we left the inner radius of the disk, the inclination and the line energy free to vary, although the line energy was constrained to range between 6.4 and 7 keV. We further assumed a disk with a single emissivity index which we fixed to~3. For the models with \textsc{Kerrdisk} and \textsc{Kyrline} we fixed the spin of the black hole to $0.998$. The results are unaffected if we fix the spin of the black hole to zero. As determined from these relativistic line profiles, the inclination of the disk is $89.9^{\circ}$ (with a 95\% confidence lower limit of $69.7^{\circ}$), $90^{\circ}$ (with a 95\% confidence lower limit of $89^{\circ}$), and $77.6^{\circ}$, (with the 95\% confidence limits of $64.6^{\circ}$ and $86.1^{\circ}$), for the \textsc{diskline}, \textsc{Kerrdisk}, and \textsc{Kyrline} profiles, respectively. For comparison, for the \textsc{Laor} profile we found a 95\% confidence lower limit for the inclination of $\sim$84$^{\circ}$ (see Tab.~\ref{tab:more models}).

The value of the inner disk radius inferred from the line profile also depends somewhat on the model that we used to describe the line. For instance, the \textsc{pl+diskbb+Laor} model gives $R_{\rm in}$ = 14.6 R$_{\rm g}$, with a 95\% confidence upper limit of 19.7 R$_{\rm g}$ (see Tab.~\ref{tab:more models}). For the \textsc{pl+diskbb+diskline} model we find $R_{\rm in}$ = 11.7 R$_{\rm g}$, with a 95\% confidence range of 6.2-15.5 R$_{\rm g}$, whereas the \textsc{pl+diskbb+Kerrdisk} and the \textsc{pl+diskbb+Kyrline} model give $R_{\rm in}$ = 1.9 R$_{\rm g}$, with a 95\% confidence range of 1.7--3.8 R$_{\rm g}$, and $R_{\rm in}$ = 15.5 R$_{\rm g}$, with a 95\% confidence range of 11.8--20.1 R$_{\rm g}$, respectively.

We note that we used here the same data as those used by M06, and our analysis is similar to theirs; while here we report on a broad iron emission line with an equivalent width of 90 eV. Using the same continuum model, M06 reported an upper limit of 60 eV for the Fe line. We repeated the analysis of Swift J1753 where we applied the same spectral resolution used by M06 (a minimum of 10 counts per bin) and we simultaneously fit the PN, PCA and HEXTE spectra with a \textsc{diskbb+pl} model affected by absorption. We found best-fit parameters and reduced $\chi^{2}$ values consistent with those in M06. We then added a \textsc{Laor} component to the model with the same parameter settings as described in \S\ref{sec:general fit}. The \textsc{diskbb+pl+Laor} model gives $\chi^{2}$/$\nu$ = 2003/1951 with the line parameters consistent with those that we found. The line has a normalization that is more than 6$\sigma$ different from zero, and an equivalent width of 73.5 eV which is, within the confidence limits, consistent with the 90 eV measurement that we report.
\section{Discussion}\label{sec:discussion}
We analyzed simultaneous observations of Swift J1753 in the low/hard state taken with \textit{XMM-Newton} and \textit{RXTE} in March 2006. At variance with the results of Miller et al. (2006a), we can successfully fit the X-ray spectrum of Swift J1753 with continuum models that do not require emission from a prominent disk extending down to the innermost stable circular orbit. Models including emission from a disk component, as proposed by M06, are also consistent with the data, although for those models where the disk component is required we found that the flux of the disk is a factor of 2 to 3 less than in the analysis of M06. Regardless of the continuum model, we found a broad iron line at around 7 keV with an equivalent width of 60--187 eV, depending on the model. If the line is from a single ionization stage of iron and the broadening is due to relativistic effects near the black hole, then the line profile implies an inner disk radius between $\sim$2 R$_{\rm g}$ and $\sim$16 R$_{\rm g}$, depending on the relativistic line profile (i.e. Laor, diskline, Kerrdisk or Kyrline), or the model used to fit the continuum. The inner disk radius can be larger if the width of the line results from the combination of lines from different ionization stages of iron, with each of these lines being broadened by relativistic effects. In fact, a model of reflection from an ionized disk that calculates the strength of all relevant lines self-consistently, convolved with a kernel that accounts for the relativistic effects close to the black hole yields an inner disk radius of 256 R$_{\rm g}$, with a 95\% confidence lower limit of 246 R$_{\rm g}$.

To get a better insight in the accretion disk geometry there are several factors that must be considered. We know that black-hole binaries show two basic states, the low/hard and the high/soft states (see e.g. Tananbaum et al. 1972), with transitions in between these states (see e.g., McClintock \& Remillard 2006; Homan \& Belloni 2005). The X-ray emission in the HSS is dominated by a thermal component which is usually identified as emission from the accretion disk. In this state, the inner radius of the disk, $R_{\rm in}$, appears to remain constant despite large changes of the disk flux (e.g., Tanaka \& Lewin 1995; M\'{e}ndez, Belloni \& van der Klis 1998). Tanaka (1992) interpreted this lack of change in $R_{\rm in}$ as a signature of the innermost stable orbit around a black hole. The black hole spectra in the LHS is dominated by emission from a power-law like component; in those cases where the interstellar absorption towards the source is low, a soft component with a low temperature ($\lesssim$0.5 keV) appears to be present in the spectrum (e.g., XTE J1118+480; McClintock et al. 2001). This component has been interpreted as due to a cool accretion disk truncated at a large radius (e.g., Chaty et al. 2003). This picture is strengthened by the fact that during transitions from the HSS to the LHS the radius of the disk increases with a decreasing disk temperature (Gierli\'{n}ski et al. 2008), suggesting that the disk recedes and cools down during the transition. Other sources show a similar behavior (e.g., Kalemci et al. 2004), but since the inferred temperatures of the disk are below $\sim$0.3--0.5 keV, interstellar absorption and inadequate coverage of the low-energy part of the spectrum make some of those results less secure. Changes of characteristic frequencies in the power density spectra of BHCs during state transitions (e.g., M\'{e}ndez \& van der Klis 1997; Homan \& Belloni 2005) also seem to support the idea of a receding disk, although in this case there is no clear-cut explanation about the origin of the variability (see the discussion in M06), and hence the evidence is less compelling.

The picture of an accretion disk truncated at large radii in the LHS is also expected in some models. For instance, in the advection-dominated accretion-flow (ADAF) model in which the accretion flow is divided into two zones: A geometrically thin, optically thick accretion disk with a large inner radius, and a vertically extended inner region, the ADAF region, which is hot and optically thin and radiates less efficiently (Narayan, McClintock \& Yi 1996; Esin et al. 1997; see Narayan \& McClintock 2008 for a review). This model has been successfully applied not only to sources in the LHS, but to quiescent sources as well (Narayan et al. 1996). Some recent work (Meyer et al. 2007; Liu et al. 2007) suggests, however, that a part of the ADAF material may recondense back into a disk, but the contribution of this condensed disk to the total X-ray luminosity would be small (Taam et al. 2008). 

In the last five years, \textit{XMM-Newton} and \textit{Chandra} revealed relatively broad iron emission lines of BHCs with profiles that are consistent with being broadened by relativistic effects near the black hole (see Miller 2007 for a review). These broad iron lines are generally detected in the HSS (e.g., GRO J1655$-$40; Ba{\l}uci\'{n}ska-Church \& Church 2000; XTE J1650$-$500; Miller et al. 2002b; GX 339$-$4; Miller et al. 2004; Dunn et al. 2008), in which the disk is thought to extend down to the ISCO (as described before), and the relativistic effects are therefore strong. Broad iron emission lines, consistent with a disk extending close to the ISCO, were also detected in GX 339$-$4 in the LHS (Miller et al. 2006b), and in V4641 Sgr when the source was in an intermediate state (Miller et al. 2002a). We found a broad iron line in the \textit{XMM-Newton/RXTE} spectrum of Swift J1753, making it only the second case of a BHC in the LHS state showing such a broad line. A relativistic broadened iron line in this black hole state would be in contradiction with the standard scenario as described above, in which the disk is truncated at very large distances from the black hole. 

This contradiction appears to be resolved by recent studies of the BHCs Swift J1753, GX 339$-$4 and XTE J1817$-$330 (M06; Miller et al. 2006b; Rykoff et al. 2007) that reveal the possible existence of soft emission below $\sim$2 keV in these systems. This emission has been interpreted by these authors in terms of a prominent accretion disk in the LHS, with temperatures of 0.2--0.3 keV, and inner disk radii of $\sim$2--6 R$_{\rm g}$. The idea of a cool disk extending all the way down to the ISCO in the LHS has recently been challenged by Gierli\'{n}ski et al. (2008) for the case of XTE J1817$-$330 and by D'Angelo et al. (2008) for the cases of Swift J1753 and GX 339$-$4. These authors pointed out that a truncated disk would be affected by irradiation from the same Comptonized photons that produce the high-energy emission, and would therefore appear to be hotter and a have smaller radius than in the non-irradiated case. This effect was not taken into account in the fits of M06, Miller et al. (2006b), or Rykoff et al. (2007). Gierli\'{n}ski et al. (2008) fit the \textit{Swift} spectra of XTE J1817$-$330 with a simplified model that accounts for irradiation, and they find that if this effect is not considered, the inner disk radius is underestimated by a factor of 2--3. D'Angelo et al. (2008) analyzed the case of an ADAF flow inside a truncated accretion disk and, from qualitative fits to the X-ray spectra of Swift J1753 and GX 339$-$4, they conclude that reprocessing of hard photons by a truncated disk would mimic a soft disk extending down to very small radii. 

Using the same data as analyzed by M06, we found that there are realistic models that fit the X-ray spectrum of Swift J1753 without the need of a thermal disk-like component extending down to very small radii. We point out that for our fits we do not consider the effect of possible irradiation of the disk by high-energy Comptonized photons (Gierli\'{n}ski et al. 2008; D'Angelo et al. 2008). We can calculate the magnitude of this effect by noting that in the case of the \textsc{reflion+pl} model only $\sim$30\% of the power-law flux is reflected off the accretion disk, and therefore the remaining $\sim$70\% of this flux must be thermalized and re-emitted by the disk. From our results we deduced that the disk flux due to this irradiation would have to be $\sim$2 $\times 10^{-10}$ erg cm$^{-2}$ s$^{-1}$ (0.6--10 keV). However, in the fit with the \textsc{reflion+pl} model we did not detect emission from the accretion disk, with an upper limit that is a factor $\sim$25 less than what is expected. (We note that the discrepancy is even larger given that one should also include the intrinsic -gravitational- disk flux; see Gierli{\'n}ski et al. 2008.) This a challenge to the interpretation of the Fe emission line as due to reflection.

In the case of GX 339$-$4, Tomsick et al. (2008) found that the disk component is significantly required to fit the LHS spectra at luminosities of 2.3\% and 0.8\% of the Eddington luminosity, a factor of $\sim$2 and $\sim$7 below the luminosities as probed by Miller et al. (2006b) for the same source. Tomsick et al. (2008) found inner disk radii of $\sim$4 R$_{\rm g}$ however, following the results of Gierli\'{n}ski et al. (2008) and D'Angelo et al. (2008), their values would probably increase by a factor of $\sim$2--3 if irradiation of the disk by the Comptonized emission was considered. Tomsick et al. (2008) also detected broad features due to iron K$\alpha$ in the LHS of GX 339$-$4; their results indicate that if the width of the line is produced by relativistic effects in the disk, then the line must originate from within 10 R$_{\rm g}$ of the black hole.

In this work we used relativistic line profiles to fit the broad emission line at around 7 keV, although one should bear in mind that there are alternative models that explain the width and the profile of the iron K$\alpha$ line without the need of relativistic effects. For instance, Laurent \& Titarchuk (2007) proposed a model that explain the properties of the iron line in terms of down-scattering of hard photons in a Comptonizing outflow with optical depth greater than 1. Done \& Gierli\'{n}ski (2006) applied a similar model to fit \textit{BeppoSAX} data of the galactic BHC XTE J1650$-$500. They found that the line at around 7 keV in the spectrum of a bright LHS observation can either be fit with a model including extreme relativistic effects, or by resonance iron K-line absorption from an outflowing disk wind and an emission line that is compatible with a truncated disc. On the other hand, R\'{o}\.{z}a\'{n}ska \& Madej (2008) calculated atmosphere models for an accretion disk around a super-massive black hole irradiated by a hard X-ray power law, and they found that the observed spectrum contains a Compton shoulder that can contribute to the asymmetry and equivalent width of some observed Fe K$\alpha$ lines in active galactic nuclei. One should also take into account that, as we have shown here, the best-fit profile of the line depends on the underlying continuum assumed to fit the data, which adds extra uncertainties to the inferred parameters of the disk.

\section{Conclusions}\label{sec:conclusion}
We have shown that the X-ray spectrum of Swift J1753 can be fit with a continuum model that does not require a disk-like component, and inferred from the line profile the disk does not extend down to the ISCO, but it is consistent with a disk truncated at a few to a few hundred gravitational radii. However, as demonstrated by M06, the data of Swift J1753 also allow for fits with models in which the disk is truncated at radii close to the ISCO, although irradiation of the disk by the Comptonized emission may affect this picture (Gierli\'{n}ski et al. 2008; D'Angelo et al. 2008). We found that in those cases that a disk-like component was required, the contribution of the disk to the total flux is a factor of 2 to 3 times less than the flux contribution that M06 found for the disk component.

We have detected a broad iron line in the spectrum of Swift J1753. The inner radius of the disk deduced from fits to the line, using a Laor profile, seems to suggest a disk extending down to 5.5--12.5 R$_{\rm g}$, for the best-fit models. Fits with different relativistic line profiles suggest somewhat larger inner disk radii, up to $\sim$16 R$_{\rm g}$ (diskline and Kyline), or lower $R_{\rm in}$, down to $\sim$2 R$_{\rm g}$ (Kerrdisk), whereas reflection models smeared by relativistic effects suggest even much larger radii, $R_{\rm in}\sim$250 R$_{\rm g}$. The \textit{XMM-Newton/RXTE} data as analyzed here, do not provide a definitive answer to the question of the accretion disk is truncated at large radii or not, since the answer strongly depends on the continuum and line model that are used. One should however bear in mind that the interpretation of the line profile in terms of relativistic effects close to the black hole is not the only possible one, and other alternatives do not require a disk extending down close to the ISCO to explain the shape of the line. It remains to be seen whether this kind of models can fit the data of Swift J1753 and other sources in the LHS.
\section*{Acknowledgments}
This work is based on observations obtained with \textit{XMM-Newton}, an ESA science mission with instruments and contributions directly funded by ESA Member States and the USA (NASA). We thank Diego Altamirano for useful comments and discussions, and Matteo Guainazzi for providing the pharbn tool. We are grateful to the \textit{XMM-Newton} help desk for their assistance and advise. TMB acknowledges support from the International Space Science Institute (ISSI) and from ASI via contract I/088/06/0.

\bsp

\label{lastpage}


\clearpage
\pagestyle{empty}

\vspace{-0.5cm}
\setcounter{table}{0}
\begin{landscape}
\begin{table}
\begin{center}{\scriptsize
\caption{Results of various continuum models. %
In the first part of the table we show the results of a power-law (\textsc{pl}) based model. A line component (\textsc{Laor}), a disk blackbody (\textsc{diskbb}) or both were added (these components are also added to the other continuum models). The \textsc{pl} parameters are the photon index, $\Gamma$, and the normalization $N_{\rm \Gamma}$ (in units of photons/keV/cm$^{2}$/s at 1 keV). The \textsc{Laor} parameters are the line energy, $E_{\rm L}$, which we constrain to range between 6.4 and 7 keV, the emissivity index, $index$, inner disk radius, $R_{\rm in}$ (with R$_{\rm g}$=$GM/c^{2}$), inclination, $i$, and normalization of the line, $N_{\rm L}$ (in units of photons/cm$^{2}$/s). The \textsc{diskbb} parameters are the temperature at inner disk radius, $kT_{\rm in}$, and disk normalization, $N_{\rm D}$, which is defined as [$R_{\rm in}$(km)/$d$(10 kpc)]$^{2}$cos$i$, where $d$ is the distance to the source. %
Next, we show the results of a broken power-law (\textsc{bknpl}) based model, with $\Gamma_{1}$, the photon index for energies below the break energy, break point, $E_{\rm b}$, photon index after the break, $\Gamma_{2}$, and normalization, $N_{\rm \Gamma}$ (same units as \textsc{pl} normalization). %
Next, we give the results of a continuum model based on a Comptonization (\textsc{comptt}) component. The \textsc{comptt} parameters are: soft seed photon temperature, $kT_{0}$, plasma temperature, $kT$, optical depth of the plasma, $\tau$, and normalization, $N_{\rm c}$. %
Next, we show the results of two reflection models \textsc{pexrav} and \textsc{pexriv}, with parameters: photon index, $\Gamma$, cosine of the inclination angle, cos$i$, and normalization, $N_{\rm P}$ (same units as \textsc{pl} normalization). An extra parameters of the \textsc{pexriv} component is the disk ionization parameter, $\xi$. %
Finally, we give the results of another reflection model: \textsc{reflion+pl}. To gain information about the disk radius, we convolve this model with a \textsc{Kerrconv} component.  The \textsc{reflion} components are the ionization parameter, $\xi_{i}$, and normalization of the reflected spectrum, $N_{\rm R}$. The photon index of the \textsc{reflion} component is coupled to the \textsc{pl} photon index. The free \textsc{Kerrconv} parameters are the inclination and inner radius of the disk. %
We fit each of the above models including an absorption model (\textsc{phabs}) with $N_{\rm H}$ the equivalent hydrogen column, and a constant component to normalize between the instruments. For all continuum models, the unabsorbed flux is 3.9$\times 10^{-10}$ erg cm$^{-2}$ s$^{-1}$ (0.6--10 keV). In the models with a \textsc{diskbb} component, the flux of this component ranges between 0.3--1.6 $\times$ 10$^{-11}$ erg cm$^{-2}$ s$^{-1}$ (0.6--10 keV).}
\label{tab:more models}
\begin{tabular}{lclccccccccccc}\hline\hline
\multicolumn{1}{r}{} & 
\multicolumn{2}{c}{\hrulefill~\textsc{pl}~\hrulefill} & 
\multicolumn{2}{r}{} & 
\multicolumn{5}{c}{\hrulefill~\textsc{Laor}~\hrulefill} & 
\multicolumn{2}{c}{\hrulefill~\textsc{diskbb}~\hrulefill} & 
\multicolumn{1}{c}{\hrulefill~\textsc{phabs}~\hrulefill} & \\
\multicolumn{1}{r}{Parameter} & 
\multicolumn{1}{c}{} & 
\multicolumn{1}{c}{$N_{\Gamma}$} & 
\multicolumn{2}{c}{} & 
\multicolumn{1}{c}{$E_{\rm L}$} & 
\multicolumn{1}{c}{} & 
\multicolumn{1}{c}{$R_{\rm in}$} & 
\multicolumn{1}{c}{$i$} & 
\multicolumn{1}{c}{$N_{\rm L}$} & 
\multicolumn{1}{c}{$kT_{\rm in}$} & 
\multicolumn{1}{c}{} & 
$N_{\rm H}$ & 
 \\
\multicolumn{1}{l}{Model}  & 
\multicolumn{1}{c}{$\Gamma$} & 
\multicolumn{1}{c}{($10^{-2}$)} & 
\multicolumn{2}{c}{} & 
\multicolumn{1}{c}{(keV)} & 
\multicolumn{1}{c}{$index$} & 
\multicolumn{1}{c}{(R$_{\rm g}$)} & 
\multicolumn{1}{c}{(deg)} & 
\multicolumn{1}{c}{($10^{-4}$)} & 
\multicolumn{1}{c}{(keV)} & 
\multicolumn{1}{c}{$N_{\rm D}$} & 
($10^{21}$ cm$^{-2}$) & 
$\chi^{2}$/$\nu$ \\  \hline
\textsc{pl+diskbb}   & 
1.60$\pm$0.01 & 
5.57$\pm$0.05 &
-- & -- & -- & -- & -- & -- & -- & 
0.38$\pm$0.04 & 
36.7$\pm$10.4 & 
1.61$\pm$0.02 & 
603.7/500\\ 
\textsc{pl+Laor} & 
1.66$\pm$0.01 & 
6.05$\pm$0.02 &
-- & -- &
6.40$^{+0.16}_{-0.0}$ & 
3.8$\pm$0.5 & 
1.36$^{+0.31}_{-0.13}$ & 
90.0$^{+0.0}_{-1.8}$ & 
5.8$\pm$0.7 & 
-- & -- & 
1.60$\pm$0.01 & 
646.1/497\\
\textsc{pl+Laor+diskbb}	  & 
1.62$\pm$0.01 & 
5.71$\pm$0.04 &
-- & -- &
6.79$\pm$0.17 & 
3.9$^{+0.4}_{-0.9}$ & 
14.6$\pm$5.1 & 
86.2$^{+0.1}_{-2.3}$ & 
1.8$\pm$0.3 & 
0.36$\pm$0.02 &  
35.2$^{+3.2}_{-2.7}$ & 
1.63$\pm$0.02 & 
564.2/495\\\hline
%
%
\multicolumn{1}{r}{} & 
\multicolumn{4}{c}{\hrulefill~\textsc{bknpl}~\hrulefill} & 
\multicolumn{5}{c}{\hrulefill~\textsc{Laor}~\hrulefill} & 
\multicolumn{2}{c}{\hrulefill~\textsc{diskbb}~\hrulefill} & 
\multicolumn{1}{c}{\hrulefill~\textsc{phabs}~\hrulefill} & \\
\multicolumn{1}{r}{} & 
\multicolumn{1}{c}{} & 
\multicolumn{1}{c}{$E_{\rm b}$} & 
\multicolumn{1}{c}{} & 
\multicolumn{1}{c}{$N_{\Gamma}$} & 
\multicolumn{1}{c}{$E_{\rm L}$} & 
\multicolumn{1}{c}{} & 
\multicolumn{1}{c}{$R_{\rm in}$} & 
\multicolumn{1}{c}{$i$} & 
\multicolumn{1}{c}{$N_{\rm L}$} & 
\multicolumn{1}{c}{$kT_{\rm in}$} & 
\multicolumn{1}{c}{} & 
$N_{\rm H}$ &   \\
\multicolumn{1}{l}{}  & 
\multicolumn{1}{c}{$\Gamma_{1}$} &
\multicolumn{1}{c}{(keV)} &
\multicolumn{1}{c}{$\Gamma_{2}$} & 
\multicolumn{1}{c}{($10^{-2}$)} & 
\multicolumn{1}{c}{(keV)} & 
\multicolumn{1}{c}{$index$} & 
\multicolumn{1}{c}{(R$_{\rm g}$)} & 
\multicolumn{1}{c}{(deg)} & 
\multicolumn{1}{c}{($10^{-4}$)} & 
\multicolumn{1}{c}{(keV)} & 
\multicolumn{1}{c}{$N_{\rm D}$} & 
($10^{21}$ cm$^{-2}$) & 
$\chi^{2}$/$\nu$ \\  \hline
\textsc{bknpl}	  & 
1.69$\pm$0.01 & 
2.9$\pm$0.2 & 
1.60$\pm$0.01 & 
6.17$\pm$0.04 & 
-- & -- & -- & -- & -- & -- &  -- & 
1.67$\pm$0.02 & 
601.8/500\\
\textsc{bknpl+Laor} & 
1.70$\pm$0.01 & 
2.7$\pm$0.2 & 
1.62$\pm$0.01 & 
6.18$\pm$0.05 & 
6.75$^{+0.25}_{-0.12}$ & 
3.7$^{+6.3}_{-0.9}$ & 
12.5$^{+11.9}_{-1.7}$ & 
86.2$^{+0.1}_{-2.8}$ & 
1.8$\pm$0.8 & 
-- & -- & 
1.67$\pm$0.02 & 
564.0/495\\
\textsc{bknpl+Laor+diskbb}  & 
1.71$\pm$0.04 & 
2.9$\pm$0.2 & 
1.62$\pm$0.01 & 
6.36$\pm$0.34 & 
6.89$^{+0.11}_{-0.49}$ & 
3.9$^{+6.1}_{-1.2}$ & 
17.4$\pm$7.9 & 
86.2$^{+0.1}_{-11.3}$ & 
1.5$^{+0.7}_{-0.4}$ & 
0.18$\pm$0.02 &  
2458.8$^{+4484.6}_{-1908.1}$ & 
2.09$^{+0.16}_{-0.07}$ & 
552.9/493\\ \hline
%
\multicolumn{1}{r}{} & 
\multicolumn{4}{c}{\hrulefill~\textsc{comptt}~\hrulefill} & 
\multicolumn{5}{c}{\hrulefill~\textsc{Laor}~\hrulefill} & 
\multicolumn{2}{c}{\hrulefill~\textsc{diskbb}~\hrulefill} & 
\multicolumn{1}{c}{\hrulefill~\textsc{phabs}~\hrulefill} & \\
\multicolumn{1}{r}{} & 
\multicolumn{1}{c}{$kT_{0}$} & 
\multicolumn{1}{c}{$kT$} & 
\multicolumn{1}{c}{} & 
\multicolumn{1}{c}{$N_{\rm C}$} & 
\multicolumn{1}{c}{$E_{\rm L}$} & 
\multicolumn{1}{c}{} & 
\multicolumn{1}{c}{$R_{\rm in}$} & 
\multicolumn{1}{c}{$i$} & 
\multicolumn{1}{c}{$N_{\rm L}$} & 
\multicolumn{1}{c}{$kT_{\rm in}$} & 
\multicolumn{1}{c}{} & 
$N_{\rm H}$ &  
 \\
\multicolumn{1}{l}{}  & 
\multicolumn{1}{c}{(keV)} &
\multicolumn{1}{c}{(keV)} &
\multicolumn{1}{c}{$\tau$} & 
\multicolumn{1}{c}{($10^{-2}$)} & 
\multicolumn{1}{c}{(keV)} & 
\multicolumn{1}{c}{$index$} & 
\multicolumn{1}{c}{(R$_{\rm g}$)} & 
\multicolumn{1}{c}{(deg)} & 
\multicolumn{1}{c}{($10^{-4}$)} & 
\multicolumn{1}{c}{(keV)} & 
\multicolumn{1}{c}{$N_{\rm D}$} & 
($10^{21}$ cm$^{-2}$) & 
$\chi^{2}$/$\nu$ \\  \hline
\textsc{comptt+Laor}  & 
0.13$\pm$0.01 & 
11.9$\pm$2.4 & 
3.2$\pm$0.3 & 
1.7$\pm$0.2 & 
6.79$\pm$0.20 & 
3.8$^{+2.5}_{-0.9}$ & 
11.1$^{+4.9}_{-3.8}$ & 
86.2$^{+1.1}_{-0.1}$ & 
2.7$\pm$0.6 & 
-- &  
-- & 
1.55$\pm$0.03 & 
662.8/495\\
\textsc{comptt+Laor+diskbb}  & 
0.01$^{+0.05}_{-0.01}$ & 
23.1$^{+3.0}_{-2.0}$ & 
2.2$^{+0.5}_{-0.03}$ & 
4.1$^{+0.1}_{-0.8}$ & 
6.75$^{+0.25}_{-0.35}$ & 
3.5$^{+6.5}_{-0.7}$ & 
12.4$^{+11.6}_{-7.4}$ & 
86.2$^{+0.1}_{-2.8}$ & 
1.6$\pm$0.6 & 
0.38$\pm$0.04 &  
35.1$^{+29.9}_{-17.3}$ & 
1.60$\pm$0.04 & 
566.4/493\\ \hline
\multicolumn{1}{r}{} & 
\multicolumn{4}{c}{\hrulefill~\textsc{pexrav/pexriv}~\hrulefill} & 
\multicolumn{5}{c}{\hrulefill~\textsc{Laor}~\hrulefill} & 
\multicolumn{2}{r}{\hrulefill~\textsc{diskbb}~\hrulefill} & 
\multicolumn{1}{c}{\hrulefill~\textsc{phabs}~\hrulefill} & \\
\multicolumn{1}{r}{} & 
\multicolumn{1}{c}{} & 
\multicolumn{1}{c}{} & 
\multicolumn{1}{c}{$\xi$} & 
\multicolumn{1}{c}{$N_{\rm P}$} & 
\multicolumn{1}{c}{$E_{\rm L}$} & 
\multicolumn{1}{c}{} & 
\multicolumn{1}{c}{$R_{\rm in}$} & 
\multicolumn{1}{c}{$i$} & 
\multicolumn{1}{c}{$N_{\rm L}$} & 
\multicolumn{1}{c}{$kT_{\rm in}$} &
\multicolumn{1}{c}{} &
$N_{\rm H}$ &  \\
\multicolumn{1}{l}{}  & 
\multicolumn{1}{c}{$\Gamma$} &
\multicolumn{1}{c}{cos$i$} &
\multicolumn{1}{c}{(erg cm/s)} & 
\multicolumn{1}{c}{($10^{-2}$)} & 
\multicolumn{1}{c}{(keV)} & 
\multicolumn{1}{c}{$index$} & 
\multicolumn{1}{c}{(R$_{\rm g}$)} & 
\multicolumn{1}{c}{(deg)} & 
\multicolumn{1}{c}{($10^{-4}$)} & 
\multicolumn{1}{c}{(keV)} &
\multicolumn{1}{c}{$N_{\rm D}$} &
($10^{21}$ cm$^{-2}$) & 
$\chi^{2}$/$\nu$ \\  \hline
\textsc{pexrav}  & 
1.67$\pm$0.01 & 
0.16$\pm$0.04 & 
-- & 
6.06$\pm$0.03 &
-- & -- &-- & -- & -- & -- & -- & 
1.61$\pm$0.02 & 
693.7/501\\
\textsc{pexrav+Laor}  & 
1.68$\pm$0.01 & 
0.13$\pm$0.05 & 
-- & 
6.11$\pm$0.04 &
6.40$^{+0.12}_{-0.0}$ & 
3.4$^{+0.7}_{-0.4}$ &
5.5$\pm$0.8 & 
86.9$\pm$2.7 & 
3.4$\pm$0.6 & 
-- & 
-- & 
1.64$\pm$0.02 & 
590.0/496\\
\textsc{pexrav+Laor+diskbb}  & 
1.65$\pm$0.01 & 
0.01$^{+0.05}_{-0.0}$ & 
-- & 
5.90$\pm$0.07 &
6.40$^{+0.12}_{-0.0}$ & 
3.2$\pm$0.5 &
5.6$^{+1.5}_{-0.9}$ & 
84.8$^{+2.4}_{-10.8}$ & 
2.5$^{+0.6}_{-1.3}$ & 
0.31$\pm$0.05 & 
50.7$^{+43.7}_{-32.4}$ & 
1.68$\pm$0.06 & 
560.9/494\\
\textsc{pexriv} & 
1.68$\pm$0.01    & 
0.13$\pm$0.03    & 
134$^{+113}_{-74}$ & 
6.09$\pm$0.03    &
-- & -- &-- & -- & -- & -- & -- & 
1.64$\pm$0.02 & 
651.6/500\\
\textsc{pexriv+Laor} & 
1.69$\pm$0.01 & 
0.12$\pm$0.03 & 
0.0$^{+0.01}$ & 
6.12$\pm$0.03 &
6.40$^{+0.1}_{-0.0}$ & 
3.4$\pm$0.6 &
5.5$\pm$0.7 & 
84.8$^{+1.4}_{-2.3}$ & 
3.5$\pm$0.7 & 
-- &
-- &
1.65$\pm$0.02 & 
586.4/495\\
\textsc{pexriv+Laor+diskbb} & 
1.64$\pm$0.01 & 
0.01$^{+0.05}_{-0.0}$ & 
317$^{+118}_{-205}$ & 
5.85$\pm$0.07 &
6.43$^{+0.57}_{-0.03}$ & 
10.0$^{+0.0}_{-6.2}$ &
5.6$\pm$3.2 & 
78.0$^{+8.3}_{-18.2}$ & 
1.0$^{+0.9}_{-0.4}$ & 
0.35$\pm$0.03 &
31.0$^{+10.1}_{-7.2}$ &
1.67$\pm$0.04 & 
562.2/493\\ \hline
\multicolumn{1}{c}{} &
\multicolumn{2}{c}{\hrulefill~\textsc{pl}~\hrulefill} &
\multicolumn{1}{c}{} &
\multicolumn{3}{c}{\hrulefill~\textsc{reflion}~\hrulefill} &
\multicolumn{1}{c}{} &
\multicolumn{2}{c}{\hrulefill~\textsc{Kerrconv}~\hrulefill} &
\multicolumn{2}{c}{\hrulefill~\textsc{diskbb}~\hrulefill} &
\multicolumn{1}{c}{\hrulefill~\textsc{phabs}~\hrulefill} &
\multicolumn{1}{c}{} \\
\multicolumn{2}{c}{} &
\multicolumn{1}{c}{$N_{\Gamma}$} &
\multicolumn{2}{c}{} &
\multicolumn{1}{c}{$\xi_{i}$} &
\multicolumn{1}{c}{$N_{\rm R}$} &
\multicolumn{1}{c}{} &
\multicolumn{1}{c}{$i$} &
\multicolumn{1}{c}{$R_{\rm in}$} &
\multicolumn{1}{c}{$kT_{\rm in}$} &
\multicolumn{1}{c}{} &
\multicolumn{1}{c}{$N_{\rm H}$} &
\multicolumn{1}{c}{} \\
\multicolumn{1}{c}{} &
\multicolumn{1}{c}{$\Gamma$} &
\multicolumn{1}{c}{(10$^{-2}$)} &
 \multicolumn{1}{c}{} &
\multicolumn{1}{c}{$\Gamma$} &
\multicolumn{1}{c}{(erg cm/s)} &
\multicolumn{1}{c}{(10$^{-8}$)} &
\multicolumn{1}{c}{} &
\multicolumn{1}{c}{(deg)} & 
\multicolumn{1}{c}{(R$_{\rm g}$)} &
\multicolumn{1}{c}{(keV)} &
\multicolumn{1}{c}{$N_{\rm D}$} &
\multicolumn{1}{c}{(10$^{21}$ cm$^{-2}$)} &
\multicolumn{1}{c}{$\chi^{2}/\nu$}\\\hline
\multicolumn{1}{l}{\textsc{reflion+pl}} &
1.54$\pm$0.01 &
4.10$\pm$0.15 &
-- &
1.54$\pm$0.01 &
5053$^{+528}_{-650}$ &
7.6$\pm$0.8 &
-- &
-- &
-- &
-- &
-- &
1.75$\pm$0.01 &
568.4/500\\ 
\multicolumn{1}{l}{\textsc{reflion+pl+diskbb}} &
1.59$\pm$0.01 &
5.17$^{+0.24}_{-0.09}$ &
-- &
1.59$\pm$0.01 &
1850$^{+576}_{-628}$ &
8.2$\pm$0.2 &
-- &
-- &
-- &
0.39$\pm$0.06 &
17.7$^{+18.8}_{-11.0}$ &
1.66$\pm$0.01 &
560.3/498\\ 
\multicolumn{1}{l}{\textsc{(reflion+pl)$\ast$kerrconv}} &
1.55$\pm$0.01 &
4.18$\pm$0.26 &
-- &
1.55$\pm$0.01 &
4800$^{+233}_{-96}$ &
7.9$\pm$0.3 &
-- &
0.9$^{+7.6}_{-0.9}$ &
256$\pm$12 &
-- &
-- &
1.75$\pm$0.01 &
562.1/498\\ \hline
\end{tabular}
\normalsize
}\end{center}
\end{table}
\end{landscape}
\clearpage

\end{document}